\documentclass[usenatbib]{mnras}
\pdfoutput=1
\usepackage{amssymb}
\usepackage{amsfonts}
\usepackage{amsmath}
\usepackage[pdftex]{graphicx}
\usepackage{epsfig}
\usepackage{color}
\usepackage{dblfloatfix}
\usepackage{float}
\usepackage{url}
\def\twid{\mathrel{\lower.1ex\hbox{$\sim$}}}
\def\gtwid{\mathrel{\raise.3ex\hbox{$>$\kern-.75em\lower1ex\hbox{$\sim$}}}}
\def\ltwid{\mathrel{\raise.3ex\hbox{$<$\kern-.75em\lower1ex\hbox{$\sim$}}}}
\def\\{\hfil\break}

\def\hmpc{$h^{-1}$Mpc}
\def\hkpc{$h^{-1}$kpc}

\newcommand{\be}{\begin{equation}}
\newcommand{\ee}{\end{equation}}
\newcommand{\bea}{\begin{eqnarray}}
\newcommand{\eea}{\end{eqnarray}}

\newcommand{\galann}{{ML}}
\newcommand{\mufasa}{{\sc Mufasa}}

\begin{document}

\title{Painting galaxies into dark matter halos using machine learning}

\author[Agarwal, Dav\'e, \& Bassett]{\Large
Shankar Agarwal$^{1, \star}$, Romeel Dav\'e$^{1, 2, 3, 4 \flat}$, Bruce A. Bassett$^{1,4,5,6 \dagger}$ \\
$^1$African Institute for Mathematical Sciences, 6 Melrose Road, Muizenberg, 7945, Cape Town, South Africa\\
$^2$Institute for Astronomy, Royal Observatory, Edinburgh EH9 3HJ, United Kingdom\\
$^3$University of the Western Cape, Bellville, Cape Town, 7535, South Africa\\
$^4$South African Astronomical Observatory, Observatory, Cape Town, 7925, South Africa\\
$^5$SKA South Africa, The Park, Park Road, Pinelands, Cape Town 7405, South Africa\\
$^6$Department of Maths and Applied Maths, University of Cape Town, Cape Town, South Africa\\
emails: $^{\star}$agarwalshankar@aims.ac.za, $^{\flat}$romeeld@gmail.com, $^{\dagger}$bruce.a.bassett@gmail.com
}

\date{}
\maketitle

\begin{abstract}
We develop a machine learning (ML) framework to populate large dark matter-only simulations with baryonic galaxies. Our ML framework takes input halo properties including halo mass, environment, spin, and recent growth history, and outputs central galaxy and halo baryonic properties including stellar mass ($M_*$), star formation rate (SFR), metallicity ($Z$), neutral (\ion{H}{i}) and molecular ($\rm H_2$) hydrogen mass. We apply this to the \mufasa\ cosmological hydrodynamic simulation, and show that it recovers the mean trends of output quantities with halo mass highly accurately, including following the sharp drop in SFR and gas in quenched massive galaxies. However, the scatter around the mean relations is under-predicted. Examining galaxies individually, at $z=0$ the stellar mass and metallicity are accurately recovered ($\sigma\lesssim 0.2$~dex), but SFR and \ion{H}{i} show larger scatter ($\sigma\gtrsim 0.3$~dex); these values improve somewhat at $z=1,2$. Remarkably, \galann\ quantitatively recovers second parameter trends in galaxy properties, e.g. that galaxies with higher gas content and lower metallicity have higher SFR at a given $M_*$. Testing various ML algorithms, we find that none perform significantly better than the others, nor does ensembling improve performance, likely because none of the algorithms reproduce the large observed scatter around the mean properties. For the random forest algorithm, we find that halo mass and nearby ($\sim 200$~kpc) environment are the most important predictive variables followed by growth history, while halo spin and $\sim$Mpc scale environment are not important. Finally we study the impact of additionally inputting key baryonic properties $M_*$, SFR and $Z$, as would be available e.g. from an equilibrium model, and show that particularly providing the SFR enables \ion{H}{i} to be recovered substantially more accurately.
\end{abstract}


\section{Introduction}
Galaxies are the primary markers by which we trace out the structure and evolution of the Universe over most of cosmic time.  Galaxies form within dark matter halos along the nodes and filaments of the Cosmic Web, thereby providing a luminous (albeit sparse) sampling of the underlying matter distribution.  Owing to this, the spatial distribution of galaxies represents one of the observational pillars that allows us to constrain the parameters of the Lambda Cold Dark Matter ($\Lambda$CDM) concordance cosmology, and particularly search for deviations from the paradigm that may signify new physics such as the nature of dark energy.

To do so, however, requires understanding how the light emitted by galaxies in various wavelengths traces the underlying matter distribution.  Thanks to impressive gravity-only (N-body) simulations, it is now straightforward to precisely relate the population of dark matter halos to the underlying matter distribution.  However, relating the luminous components of galaxies to the halo population is less well understood.  Galaxy formation involves a complex set of highly nonlinear and self-regulating physical processes spanning an enormous range of physical and temporal scales which cannot be solved ab initio at present~\citep{Somerville2015,Naab2017}.  As such, models of galaxy formation within a cosmological context typically invoke heuristic representations of the small-scale processes such as star formation and supernova feedback that govern galaxy growth.

Recent years have seen remarkable progress in modeling galaxy formation.  The pioneering Millennium Simulation~\citep{Springel2005} exceeded the 10 billion particle barrier for dark matter-only runs, with modern N-body simulations now on the verge of exceeding a trillion particles~\citep{Potter2017}.  Traditionally, to model the galaxy population, the N-body simulations are used as an input to so-called semi-analytic models (SAMs)~\citep{White1991,Kauffmann1993,Cole1994} that employ highly parameterized descriptions of galaxy evolution within halos, thus enabling direct predictions for how galaxies trace large-scale structure.  The technique of populating N-body simulations with SAMs has continued to grow in sophistication, but fundamentally SAMs rely on a simplified analytic description of galaxy formation and have a large number of free parameters.  Hence while they have demonstrated numerous successes in reproducing the observed galaxy population, it is unclear whether they provide sufficient robustness in modeling galaxy formation to be relied upon for precision cosmology.

A separate rapidly-advancing approach to modeling galaxy formation is through the use of hydrodynamic simulations.  Here, dark matter and gas are evolved concurrently, with the gas subject to additional hydrodynamic and radiative processes that enable the formation of baryonic condensations that are identifiable as galaxies.  Typically, such models still rely on heuristic prescriptions for sub-resolution processes such as star formation, but at least the dynamical influence of the baryons is handled in an ab initio manner.  The dynamical motion of inflows and outflows from galaxies, often called the baryon cycle, can be followed directly in such models, resulting in many emergent phenomena.  Recent simulations have achieved impressive levels of realism in modeling the observed galaxy population, and therefore offer a competitive approach to connecting galaxies and large-scale structure to SAMs, with many fewer assumptions.

Unfortunately, given current computing capabilities, hydrodynamic simulations require heroic efforts to achieve the Gpc$^3$-scale volumes required to constrain cosmology at a precision level~\citep{Remus2017}, and do not cover the dynamic range to robustly model the detailed buildup of baryonic galaxy components.  Utilizing hydro simulations for cosmology optimally thus still requires connecting galaxies to dark matter halos in some way.  A simple way might be to assume fitting functions for how galaxies populate halos as a function of halo mass or other properties, but such an approach requires choosing somewhat arbitrary functional forms and does not take advantage of all the correlations among galaxy properties as predicted in the hydro simulations.  Ideally, we desire a non-parametric, holistic, and flexible approach to characterize how galaxies populate dark matter halos.

Machine learning provides a potentially powerful approach to address this problem in a way that can exploit hidden insights and correlations in order to accurately populate galaxies into halos. Machine learning is an approach to solving problems that exploits the ability of certain algorithms to cluster data (an example of so-called ``unsupervised learning") or to successfully extrapolate from training data -- where a target quantity is known -- to test data where the target of interest is unknown (so-called ``supervised learning"). In general the target of interest may be discrete, in which case one typically considers it a classification problem (such as predicting whether a galaxy will be spiral or elliptical), or may be continuous, in which case it is known as a regression problem. Recent progress in machine learning and artificial intelligence has been spectacular, particularly in areas related to computer vision.  In astronomy and cosmology, machine learning has been applied in a myriad of ways\footnote{see http://arxiv.org/find/astro-ph/1/abs:+AND+machine+learning/0/1/0/all/0/1} including the classification of transients. See {\tt http://www.astroml.org/} for more examples and sample code.  Our problem of populating galaxies into halos is a regression problem since the output properties are continuous, and can be solved with a supervised learning approach since a known training sample exists.

In this paper, we use machine learning (ML) to ``teach" the computer how to populate galaxies into dark matter halos.  For this exercise, we train the ML on the recent \mufasa\ simulations, which have been shown to provide an impressive match to a wide range of galaxy stellar, gas, and metal properties over cosmic time~\citep{Dave2016,Dave2017a,Dave2017b}, and then test its efficacy on the separate testing subsample.  We show that using information about current halo properties and recent growth history, we can recover the stellar, neutral hydrogen, star formation rate, and metal properties of galaxies to within well-characterized uncertainties.  These physical properties directly relate to the luminous output of galaxies in various bands, providing an avenue to connect large-scale halo catalogs with observable galaxy properties in optical/near-infrared bands, nebular emission lines, and 21~cm emission.  This work thus sets the stage for machine learning to provide a competitive approach to populating large-scale N-body simulations with galaxies for cosmology applications.

The approach of using machine learning to populate dark matter halos by training on hydrodynamic simulations has previously been examined by~\citet{Kamdar2016}, who used the Illustris simulation~\citep{Vogelsberger2014} as the training set.  Our approach will be generally similar to theirs, and in areas of overlap our results quantitatively corroborate theirs.  We then go on to extend this approach in several ways: Besides utilizing \mufasa\ which agrees better with data than Illustris for key galaxy properties (particularly massive quenched systems), we specifically examine \ion{H}{i} with an eye towards future \ion{H}{i} intensity mapping surveys, we investigate the nature of the scatter by examining second parameter correlations as predicted by ML, and we explore ways to make the ML more accurate through additional inputs.  Our results lend further support to the idea that machine learning provides a valuable approach towards characterizing the baryonic and observable properties of galaxies within large-volume N-body simulations.

In \S\ref{sec:SIM} we briefly review the \mufasa\ simulation.  In \S\ref{sec:ML} we describe the input and output parameters for our ML framework, and describe the various ML methods that we employ here.  In \S\ref{sec:RESULTS} we describe our results for how well machine learning can recover intrinsic galaxy baryonic properties from only the dark matter halo catalog and its recent growth history. In \S\ref{sec:BARYONIC_INPUT} we show how \ion{H}{i} and $\rm H_2$ can be recovered more accurately when $M_*$, SFR, and $Z$ are used as input properties for \galann. In \S\ref{sec:COMPARE_ML} we compare results from the various ML algorithms explored in this paper.  Finally, we summarize and conclude in \S\ref{sec:DISCUSS}.

\section{Cosmological simulations} \label{sec:SIM}

For this work we employ the \mufasa\ cosmological hydrodynamic simulation, described fully in~\citet{Dave2016}. \mufasa\ uses the {\sc Gizmo} code~\citep{Hopkins2015} to evolve dark matter using a tree-particle-mesh scheme, and gas using a meshless finite mass (MFM) algorithm that offers important numerical advantages over traditional smoothed particle hydrodynamics solvers in terms of accuracy and robustness.

\mufasa\ includes a wide range of physical phenomena for producing realistic galaxies, including radiative cooling with metal lines, photoionisation heating, molecular gas-based star formation using a subgrid model for computing the molecular gas density~\citep{Krumholz2009}, chemical enrichment from Type II and Type Ia supernovae along with stellar evolution, and prescriptions for feedback from star formation and for quenching massive galaxies.  The star formation feedback is done kinetically using two-phase outflows, based on the results from high-resolution simulations of the Feedback in Realistic Environments project~\citep{Muratov2015}.  The quenching feedback is crudely modeled by keeping hot gas in massive halos hot as in ``radio mode" feedback~\citep{Croton2006}, which as shown by~\citet{Gabor2016} is effective at producing quenched galaxies with a sharp turnover in the galaxy mass function and a red sequence that is in reasonable agreement with observations.

\mufasa\ does comparably well or better at reproducing galaxy survey observations over a wide range of epochs, as compared to other modern hydrodynamic simulations such as Illustris~\citep{Genel2014} or Eagle~\citep{Schaye2015}.  The stellar mass function, which is a crucial barometer of how galaxy growth tracks halo growth, is in very good agreement with observations from $z\sim 0\rightarrow 6$~\citep{Dave2016}.  Gas and metal properties are also in good agreement~\citep{Dave2017a}, and \mufasa\ produces a bimodal galaxy population of quenched and star-forming systems broadly similar to that observed~\citep{Dave2017b}.  While non-trivial discrepancies versus observations remain, the overall success of \mufasa\ at matching observations suggests that it provides a state of the art platform to examine the connection between dark matter halos and key observable galaxy properties.

For this paper, we utilize the $50$ \hmpc\ \mufasa\ run, which has a minimum (adaptive) softening of $0.5$ \hkpc\ (comoving), a gas element mass of $1.8\times 10^7 M_\odot$, and a dark matter particle mass of $9.6\times 10^7 M_\odot$, allowing us to resolve galaxies down to a stellar mass of $5.8\times 10^8 M_\odot$; we will only use galaxies above this stellar mass in our analysis.  We have shown that the stellar~\citep{Dave2016} and gas~\citep{Dave2017a} mass functions are well-converged above this limit; the SFR function less so but still reasonable.  We employ a Planck-concordant cosmology with $\Omega_m=0.3$, $\Omega_\Lambda=0.7$, $\Omega_b=0.048$, $H_0=68$, $n=0.97$, and $\sigma_8=0.82$.  

We output 135 snapshots to $z=0$ of this simulation, with a typical time separation between snapshots at later epochs of $200-300$~Myr.  For every snapshot, we identify all galaxies using Spline Kernel Interpolative Denmax (SKID,~\citealt{Keres2005}), and we identify halos using a 3-D friends-of-friend halo finder that is native in the public {\sc yt} simulation analysis package with a linking length of 0.2 times the mean interparticle spacing. The linking length $b$ is roughly related to the spherical overdensity $\Delta$ w.r.t the background matter density through $\Delta/178 \approx (0.2/b)^3$~\citep{Cole1996, Courtin2011}. Quoted halo masses represent the total dynamical mass including baryons.  We use {\sc Caesar}\footnote{\tt http://caesar.readthedocs.io/en/latest/}, a {\sc yt} add-on package, to cross-match the galaxies and halos, and to compute a range of galaxy and halo properties for each snapshot.  Finally, we identify galaxy progenitors by finding the galaxy at the previous snapshot that holds the largest fraction of baryonic mass (gas+stars) that is now in the current SKID galaxy; and similarly for halo progenitors.  These {\sc Caesar} galaxy/halo catalogs, plus the progenitor information, represent the input data for the machine learning analysis presented here.

\section{Machine Learning Setup}\label{sec:ML}

Here we briefly describe the input and outputs parameters for our ML framework, and describe the various ML methods we will cross-compare to determine the most effective algorithm.

\subsection{ML inputs and outputs}\label{sec:ML_overview}

\begin{figure}
     \includegraphics[width= \columnwidth]{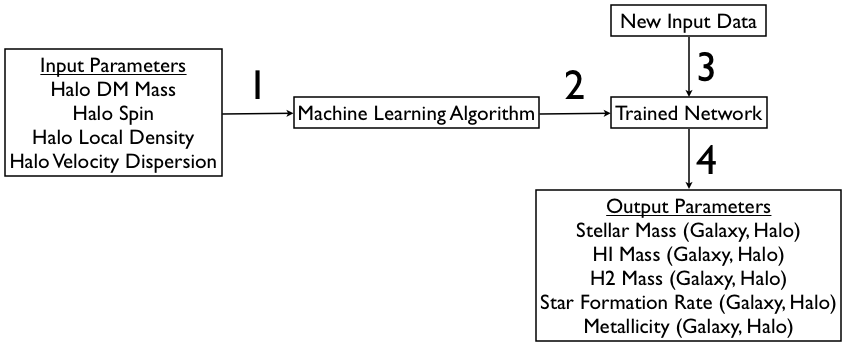}
        \caption{\small{Steps 1 and 2: a machine learning network learns to parametrize the output, for the input patterns that form the training set. Steps 3 and 4: the trained network is capable of making predictions when presented with input parameter settings. The queried input settings must lie within the parameter ranges of the patterns in the training set.
        }}
    \label{fig:ML}
\end{figure}

Fig.~\ref{fig:ML} shows a skeleton of the supervised machine learning network. Using a suitable training set (input parameters for which data is available), the machine learning algorithm is trained to learn a mapping. With this mapping the network is capable of reproducing (as closely as possible) the output, when queried with input parameter settings that are part of the training set. The trained network can now be presented with new settings of the input parameters (for which one does not have any prior data) and by using the same mapping learnt during the training process, the network makes predictions.

Adjusting the ML network parameters to get the desired mapping is called {\it training} the network. For estimating the galaxy and halo quantities, we use a training set of dark matter halos, each with the following known parameters:
\bea
\label{eq:PARAMS}
{\bf I}\equiv(\rm \{M_{\rm halo}\}_i, \{\rho\}_j, \lambda, \sigma_{vel}),
\eea
where $\{\rm M_{\rm halo}\}_i$ are the halo masses at the present snapshot ($i=0$) and the 5 immediately preceding snapshots (thus characterizing the recent merger history), $\{\rm \rho\}_j$ are the nearby halo mass densities within radii $\rm r=200, 500, 1000$ kpc centered on the halo center-of-mass, $\rm \lambda$ is the halo spin, and $\rm \sigma_{vel}$ is the halo velocity dispersion. 

The galaxy and halo quantities that \galann\ predicts are:
\bea
\label{eq:PARAMS_OUT}
{\bf O}\equiv({\rm M_*, M_{HI}, M_{H2}, SFR,} Z),
\eea
where $\rm M_{HI}$ is the \ion{H}{i} (neutral hydrogen) mass, $\rm M_{H2}$ is the molecular hydrogen mass, $M_*$ is the stellar mass, SFR is the star formation rate, and $Z$ is the SFR-weighted metallicity. We predict these 5 quantities both for entire halos and their central galaxies, thereby making $\bf O$ a 10-dimensional output space. Likewise, $\bf I$ is an 11-dimensional input space.  In this paper, we will only present central galaxy properties, for brevity; the results for the entire halos show very similar trends and do not provide significant additional insights, so we omit them and simply note that the ML algorithm is able to also predict these if required, e.g. for intensity mapping purposes.

Since the quantities in $\bf I$ and $\bf O$ span many orders of magnitude, in our fitting procedure we work with their logarithm, giving us our cost function $\chi^2$, which is presented to the ML setup for minimization with respect to the network weights $\bf w$,
\bea
\label{eq:chisq}
\chi^2({\bf w}) \!\!\!\!\!\! &=& \!\!\!\!\!\! \sum_{t=1}^T \sum_{k=1}^{N_{out}} {\rm log}^2 \left[\frac{{\bf O}^{\rm ML}(k|{\bf w,I}_t)} {{\bf O}(k|{\bf I}_t)}\right],
\eea
where the index $t$ runs over all dark matter halos in the training set, and the index $k$ runs over the $N_{out}$ quantities in ${\bf O}$ (for our case, $N_{out}=10$). ${\bf O}(k|{\bf I}_t)$ is the $k^{th}$ component (i.e., output quantity) of $\bf O$, given the input quantities ${\bf I}_t$ corresponding to the $t^{th}$ halo in the training set. The numerator is the ML prediction given the weights $\bf w$, while the denominator is the true value from \mufasa.

\subsection{ML algorithms}\label{sec:ML_DESCRIPTION}

We constructed the mapping between dark matter halo and galaxy properties using the ML routines implemented within the publicly available {\sc python} ML library, {\sc Scikit-learn}\footnote{\tt http://scikit-learn.org}~\citep{Pedregosa2012} . We explored the following ML routines: Support Vector Regression (SVR), k-Nearest Neighbors (kNN), Multilayer Perceptron (MLP), Decision Trees (DT), Random Forests (RF), Extremely Randomized Trees (ET) and AdaBoostRegressor (AdR). We briefly describe each below; for more information on each of the above ML algorithms, refer to the {\sc Scikit-learn} library.

\subsubsection{Support Vector Regression (SVR)}
SVR builds on the Support Vector Machine classification algorithm which finds the hyperplane that separates the features optimally. SVR finds the best surface to explain the data. As with SVM, one can either fit a flat hyperplane (the linear case), or via the kernel trick, fit a nonlinear function efficiently to the data. We treat the kernel as a hyper-parameter and optimize over polynomial, sigmoid and radial basis function (RBF) kernels with $C$, the error penalty parameter, varied on a grid between 0.1 and 10.  

\subsubsection{k-Nearest Neighbors (kNN)}
kNN is one of the simplest non-parametric classification and regression algorithms. It proceeds via a type of "voting" system by simply finding the $k$ nearest neighbors of the test point in feature space (where $k$ is a user-chosen integer which is a hyper-parameter that can be optimized) and using the mean of these targets to perform the regression. 

\subsubsection{Multilayer Perceptron (MLP)}
MLP (a form of an artificial neural network) is a collection of nodes arranged in a series of layers, with each node in a layer connected to all other nodes in adjacent layers. For example, a network with an architecture 5 :  15 : 10 : 4 has 5 input features, two hidden layers with 15 and 10 nodes respectively, and finally 4 output features. An extra node (called the bias node) is added to the input layer as well as to each of the hidden layers to compensate for the difference between the network's mean prediction and the mean of the outputs of training set patterns (for a review, refer \citealt{Bishop95}). The connections are referred to as MLP weights. During network training the weights are adjusted so as to minimize the difference between network predictions and the target values of the validation set patterns.

\subsubsection{Decision Trees (DT)}
DT are graphs implementing non-parametric learning. They consist of trees where each node poses a test question (e.g. "Is the metallicity greater than some threshold, $Z_*$?") to the incoming data that needs to be classified or regressed on. The answers (called leafs) either lead down the tree to new nodes or to a final prediction. The overall complexity of a DT depends on the depth of the tree which can be controlled by various parameters such as (i) maximum tree depth, (ii) minimum number of data samples to use when deciding whether to implement an internal node split, (iii) minimum error reduction (over the training set) that a daughter node must achieve over its parent node when deciding whether to implement a node split. The metric we use to select the features and thresholds at the nodes is Mean Square Error. Note that DT are susceptible to over-fitting the data, a problem mostly solved by RF.  

\subsubsection{Random Forests (RF) and Extremely Randomized Trees (ET)} RF and ET are ensemble learning methods that aggregate the predictions of many decision trees and output a mean prediction to improve  accuracy and control over-fitting.  The major difference between these two ensemble methods is that ET has many more leaf nodes, because the samples are drawn from the entire training set instead of a bootstrap sample and the splits are chosen at random from the range of values in the sample. One of the advantages of ensemble tree methods is that they automatically provide a measure of feature importance. A key feature will appear in many of the trees in the ensemble while an irrelevant feature will show up only rarely. We will use this characteristic to rank the importance of our features, as shown in Figs (\ref{fig:relative_importance}) and (\ref{fig:relative_importance_stage2}).

\subsubsection{AdaBoostRegressor (AdR)} AdR fits a base regressor (any ML routine; for example SVR) on the original dataset. A boosted ensemble is then built by fitting additional copies of the regressor on the same dataset with weights of instances adjusted according to the error of the current prediction. Dataset samples with wrong predictions get their weights boosted, thereby making subsequent regressors focus more on difficult cases. In the figures below we will show AdR's performance with the base regressors SVR and RF (which we call SVR+ and RF+), with similar results achieved for the other regressors.

It turns out that RF and SVR provide the best results on average, i.e. the tightest relationship between the training set and predictions, as we demonstrate in \S\ref{sec:COMPARE_ML}. We note here that an exhaustive search through the hyper-parameter space for each of the ML methods could improve upon our ML results, as well as change the ranking of the various ML methods. For example, using the EDEN \citep{Dufourq2017} framework to genetically evolve the optimal MLP, we were able to get MLP to perform at par with RF. However, the marginal improvement in the results suggests that we have reached the limit on the amount of information that is possible to extract from our input data.

Here we particularly focus on the RF results because RF also provides a straightforward way to assess the importance level of each input parameter in the final predictions, which is an interesting diagnostic.

\begin{figure}
     \includegraphics[width=\columnwidth]{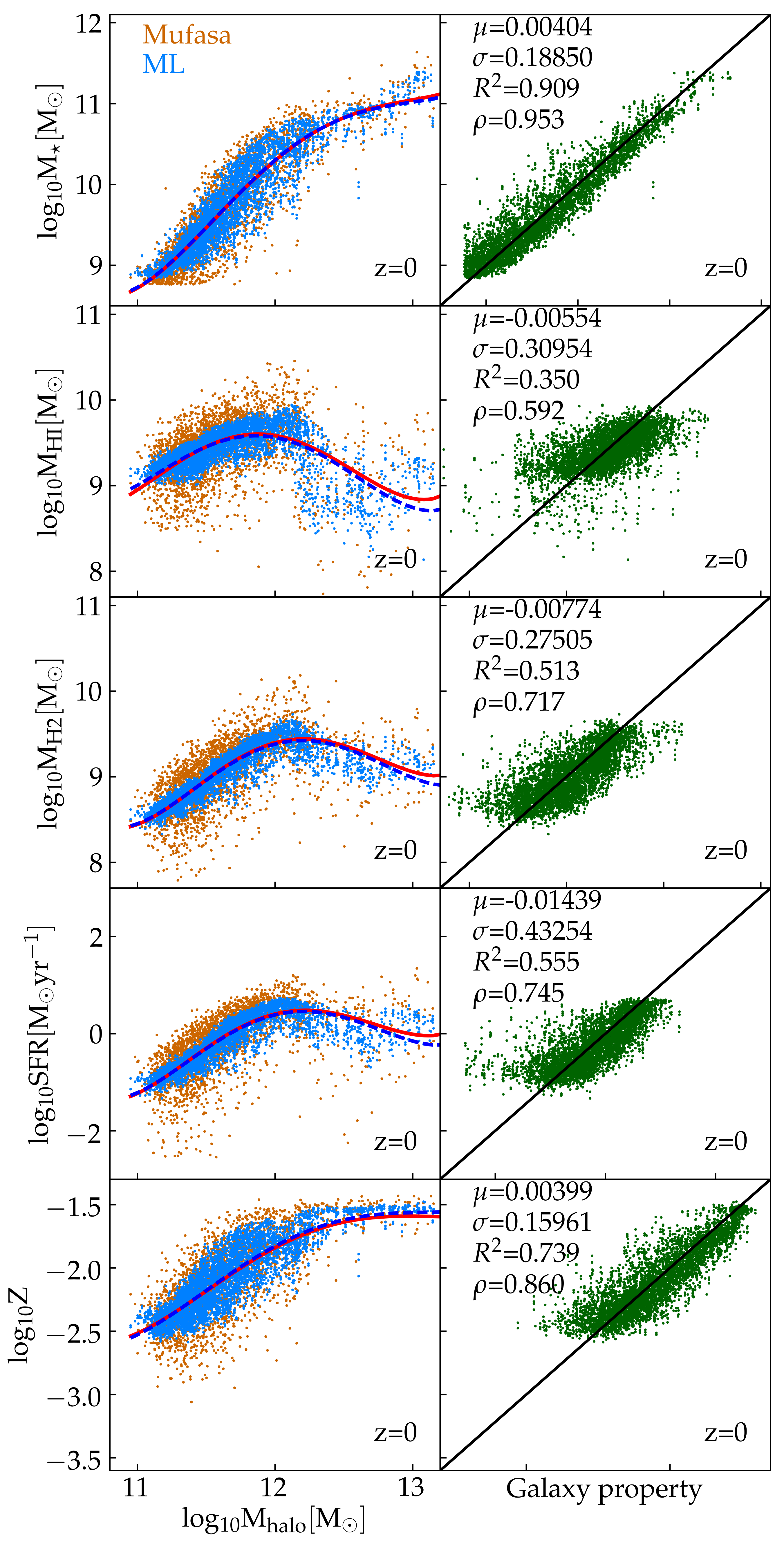}
        \caption{\small{{\it Left column:} Central galaxy properties $\rm [M_*,M_{HI},M_{H2},{\rm SFR},Z]$ vs. host halo mass, at $z=0$. The brown and blue points correspond to values from the \mufasa\ simulation and \galann\ using random forests, respectively. The mean scaling relations are shown by red solid (\mufasa) and blue dashed (\galann) lines, which mostly overlap showing that the mean trends are well recovered. The scatter in the true (\mufasa) values is however larger than predicted from \galann.  {\it Right column:} Comparison of individual central galaxy properties' true values from \mufasa\ (x-axis) vs. \galann\ predictions (y-axis).  A perfect prediction corresponds to the diagonal solid line. The mean $\mu$ and r.m.s width $\sigma$, in dex, along with the regression statistic $R^2$ and the Pearson correlation coefficient $\rho$ (see equations~\ref{eq:MU}-\ref{eq:PEARSON}) are listed in the upper left.}}
    \label{fig:only_z0}
\end{figure}

\begin{figure*}
     \includegraphics[width=\columnwidth]{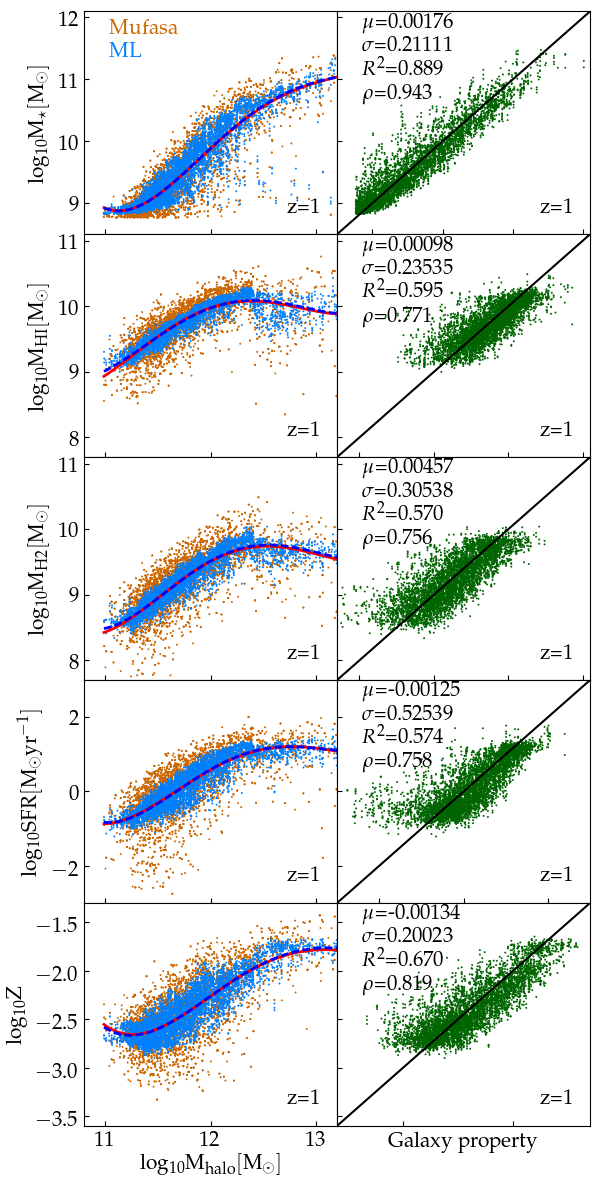}
     \includegraphics[width=\columnwidth]{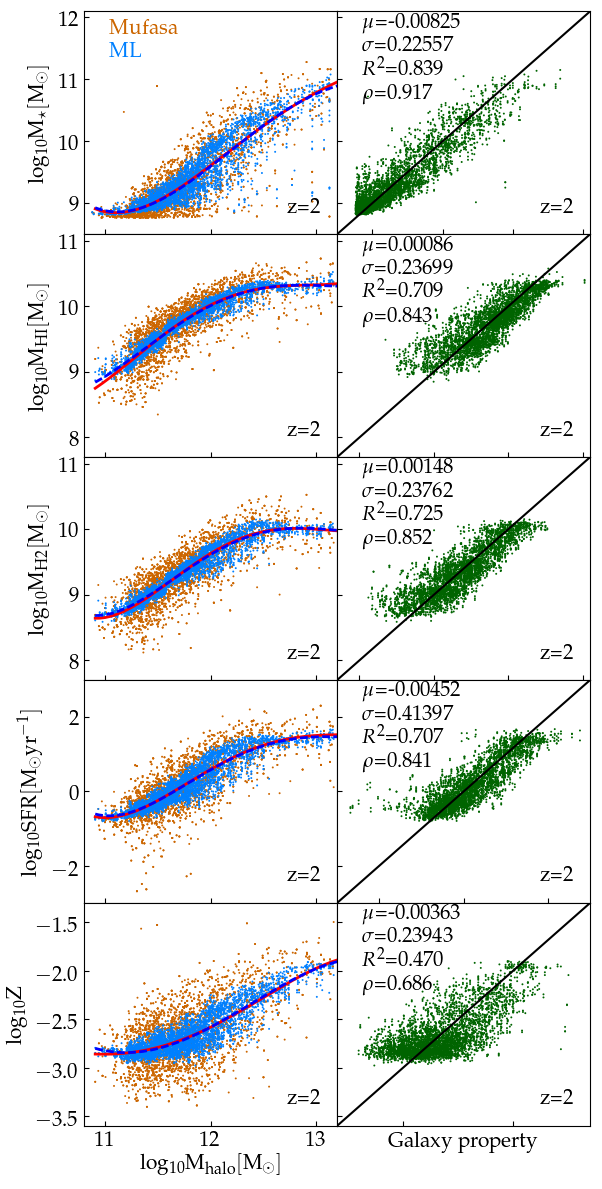}
        \caption{\small{As in Fig.~\ref{fig:only_z0} except at $z=1$ (left) and $z=2$ (right).  In each, the left panels show central galaxy properties as a function of halo mass for the original \mufasa\ simulation (brown points) and the ML prediction (blue points), and the right panels compare the ML predicted value (y-axis) vs. the true \mufasa\ values (x-axis).  At higher redshifts, the ML does a better job recovering the true values for gas and SFR than at $z=0$, but worse for $M_*$ and $Z$.}}
    \label{fig:only_z1}
\end{figure*}

\subsection{ML setup}

We divide our dataset of almost 3400 dark matter halos into training (2500), validation (500) and testing sets (400). For each halo, the 11 quantities of ${\bf I}$ are fed to the network (see Fig.~\ref{fig:ML}), while the output is held fixed at the 10 quantities of ${\bf O}$. The space of the ML hyper-parameters was randomly sampled 2000 times. The hyper-parameter configuration that minimized the cost function with respect to the halos in the validation set was chosen as the optimized network architecture. The hyper-parameters we searched for the various ML routines are the error penalty term (for SVR), the number of neighbors (for kNN), the maximum tree depth, the number of samples required to split an internal node, and the subset of input features $\in \bf I$ to consider when looking for the best split (for RF, DT and ET), the number of hidden layers and the number of nodes per hidden layer (for MLP), the number of boosts (for AdaBoostRegressor). For RF and ET, we trained a committee of 100 decision trees. 

Furthermore, in order to get an estimate of the variability in our predictions, we shuffled our deck of 3400 halos and randomly split it into training, validation and testing sets with 2500, 500 and 400 halos respectively. We repeated this 10 times, thereby giving us a set of 10 trained ML solutions.  We use these to estimate the variance in output quantities.

\section{Testing ML vs. \mufasa}\label{sec:RESULTS}

Galaxies have a variety of properties that display both overall individual trends and complex interrelationships.  Ideally, an ML algorithm would be able to learn the trends and interrelationships precisely, i.e. given {\bf I}, ML should ideally predict {\bf O} exactly (to within intrinsic uncertainties) as would be obtained from running a full hydrodynamic simulation.  In practice, however, some aspects will be easier to recover than others.  As such, we can test ML on a variety of levels.  

Here we choose to focus on testing the trends predicted by \galann\ versus halo mass $M_h$, as this is the most fundamental quantity describing halos in N-body simulations.  In this context, we can envision testing ML in four different stages, in roughly increasing order of stringency:  (i) that \galann\ reproduces the {\it mean relationship} between galaxy properties and $M_h$; (ii) that \galann\ reproduces the amplitude of the {\it scatter} around the mean relation, as a function of $M_h$; (iii) that \galann\ produces the correct galaxy property for each {\it individual} galaxy; and (iv) that \galann\ recovers intrinsic {\it higher-order correlations} between various galaxy properties.  In this section we describe how \galann\ fares in these four stages of testing using \mufasa.

Our goal is to use DM-only information to see how well the baryonic galaxy properties can be predicted.  We will show in \S\ref{sec:COMPARE_ML} that RF and SVR give the best performance, thus we choose to only show RF predictions of our five predicted quantities: neutral gas mass $\rm M_{HI}$, molecular gas mass $\rm M_{H2}$, stellar mass $M_*$, star formation rate SFR, and SFR-weighted metallicity $Z$ for halos and their central galaxies, at three redshifts $z=0,1$ and $2$.  All plots correspond to halos in the testing set, unless stated otherwise. We augment the information presented to \galann\ by providing the DM accretion history. For making predictions at $z=0$, ML is given the four halo inputs $\bf I$ (see equation~\ref{eq:PARAMS}) computed at $z=0$, as well as the halo mass, $\rm M_{\rm halo}$, from redshifts $z=0.25,0.20,0.10$ and $0.05$. Likewise, for making predictions at $z=1$, ML is given $\bf I$ computed at $z=1$, plus $\rm M_{\rm halo}$ from redshifts $z=1.5,1.4,1.3,1.2$ and $1.1$. For $z=2$, the information provided is $\bf I$ computed at $z=2$, and $\rm M_{\rm halo}$ from redshifts $z=2.5,2.4,2.3,2.2$ and $2.1$.  These choices correspond to the frequency of snapshot outputs in \mufasa.

\subsection{Mean relations and scatter}\label{sec:GalANN_vs_SIM}

Fig.~\ref{fig:only_z0} shows ML predictions using random forests for the five galaxy quantities at $z=0$. The panels are $M_*$, $\rm M_{HI}$, $\rm M_{H2}$, SFR and $Z$ (top to bottom row). The brown and blue points correspond to values from the \mufasa\ simulation and RF respectively. The mean scaling relations are shown by red solid (\mufasa) and blue dashed (\galann) lines. Note that we use a quartic polynomial fit to get the mean scaling relations. One could instead compute the means over the halo mass bins. We have checked that the two methods give similar scaling relations (except for the lowest and highest halo mass bins) and does not change our overall results.

The green panels on the right compare the RF predictions of the galaxy quantities (on y-axis) vs. their true values (on x-axis) from \mufasa. A perfect prediction corresponds to all points lying along the diagonal solid line. To quantify deviations from perfect prediction, we compute four metrics: mean deviation $\mu$, r.m.s width $\sigma$, regression score $R^2$ and the Pearson correlation coefficient $\rho$ (all listed in the right panels, with $\mu$ and $\sigma$ in dex units). If $x_i$ and $\hat{x}_i$ are the logarithms true and predicted values of the $i^{th}$ data point, then the four metrics are estimated over a sample size $N$ as
\be
\label{eq:MU}
\mu(x,\hat{x}) = \frac{1}{N} \sum_i^N (x_i - \hat{x}_i),
\ee

\be
\label{eq:SIGMA}
\sigma^2(x,\hat{x}) =  \frac{1}{N-1} \sum_i^N \left[ (x_i - \hat{x}_i) - \mu \right]^2,
\ee

\be
\label{eq:R2}
R^2(x,\hat{x}) = 1 - \frac{\sum_i^N \left[ x_i - \hat{x}_i \right]^2}{\sum_i^N \left[ x_i-<x> \right]^2},
\ee
where $<x>=\sum_i^N x_i/N$ is the mean of true values $x$.

The Pearson coefficient is a measure of the linear relationship between two quantities. In our case, for $x_i$ and $\hat{x}_i$, $\rm \rho$ estimated over a sample size $N$ is
\be
\label{eq:PEARSON}
\rho(x,\hat{x}) = \frac{\sum_i^N(x_i-<x>)(\hat{x}_i-<\hat{x}>)}{\sqrt{\sum_i^N(x_i-<x>)^2}\sqrt{\sum_i^N(\hat{x}_i-<\hat{x}>)^2}},
\ee
where $<\hat{x}>=\sum_i^N \hat{x}_i/N$ is the mean of predictions $\hat{x}$.

Comparing the ML predictions and true \mufasa\ values for the testing set at $z=0$, we see that they agree extremely well in the mean.  The red and blue mean lines in the left panels are almost completely overlapping.  This is reflected in the $\mu$ values in the right panels, which show extremely small systematic deviations typically well below 1\%.  This shows that \galann\ provides an excellent recovery of the mean trends, which is already non-trivial given that there are interesting physical features in the simulated galaxies such as the break in properties around $M_{\rm halo}\approx 10^{12}M_\odot$ above which galaxies become quenched.  We note that, owing to our imposed stellar mass resolution limit, there is a bend at the low-$M_{\rm halo}$ end of the $M_*$ curve, as well as for other quantities that correlate strongly with $M_*$; this is an artifact of the limited dynamic range of our simulation, but does not affect our conclusions.

Although the mean is well-predicted, it is clear that the scatter appears to be under-predicted, in the sense that the original \mufasa\ (brown) points span a larger range of values than the \galann\ predicted (blue) points.  This indicates that \galann\ prefers to predict objects closer to the mean relation than they should be.  The typical difference in the scatter around the mean relation is $\sim 10-30$\% between the original \mufasa\ sample and the \galann\ predictions.

This inability to fully follow the scatter manifests in deviations when comparing galaxies individually, as done in the right panels that show the original (x-axis) vs. predicted (y-axis) values for each galaxy.  Ideally, the predictions would exactly follow the 1-to-1 diagonal solid line.  While on average they do so quite well, as noted by the low values of $\mu$, there is substantial scatter around these relations that indicate \galann\ does not provide a perfect recovery.  The scatter $\sigma$ around these 1-to-1 relations, plus the other measures shown such as the regression coefficient $R^2$ and the Pearson correlation coefficient $\rho$, thus represents a measure of the accuracy of \galann.  

The scatter $\sigma$ is lowest for $M_*$ and $Z$, both less than 0.2~dex, while the gas masses $\rm M_{HI}$ and $\rm M_{H2}$ have scatter around 0.3~dex.  SFR shows the largest scatter of over 0.4 dex.  This likely owes to the fact that star formation rate is driven by processes that are fairly rapidly fluctuating, such as stochastic accretion and mergers, which is not well correlated with large-scale halo properties.  Similarly, gas content, which drives star formation, is also dependent on more rapidly varying baryonic physics.  In contrast, stellar mass and metallicity are primarily accumulated quantities, hence short-timescale variations tend to average out to yield a better correlation.

$R^2$ and $\rho$ yield qualitatively the same impression as $\sigma$: Stellar mass and metallicity show good correlation between true and predicted values, but the other quantities are worse.  In particular, \ion{H}{i} shows the poorest correlation among all the quantities.  This is particularly troubling for predicting \ion{H}{i} intensity maps, since even though the low resolution intensity mapping will Poisson-average over small-scale variations, a large predicted scatter will translate to a larger required mapping area in order to achieve the same level of statistical accuracy.

Fig.~\ref{fig:only_z1} shows the corresponding plots to Fig.~\ref{fig:only_z0} except at $z=1,2$.  The basic story is the same, in that $M_*$ and $Z$ are best-predicted, while gas content and SFR are less well predicted.  Examining the values of $\sigma$, $R^2$, and $\rho$, we see that the correlations are better for $\rm M_{HI}$, $\rm M_{H2}$, and SFR than at $z=0$, albeit slightly worse for $M_*$ and $Z$.  It appears that \galann\ does an increasingly better job at recovering these ``short-term" galaxy properties to higher redshifts, likely because quenching is less prevalent at high-$z$ and it is quenching that drives galaxies to deviate systematically from the gas and SFR values they typically have while on the galaxy main sequence.  However, ML does poorer at the ``cumulative" quantities partly owing to the smaller training sets available at these higher redshifts.

\begin{table*}
\begin{tabular}{|l|c|c|c|c|clclclcl}
 \hline
 \multicolumn{1}{l}{Property} &  \multicolumn{3}{c}{$z=0$} &  \multicolumn{3}{c}{$z=1$} &  \multicolumn{3}{c}{$z=2$} \\
 &  \multicolumn{1}{c}{$\sigma$} &  \multicolumn{1}{c}{$R^2$} &  \multicolumn{1}{c}{$\rho$} &  \multicolumn{1}{c}{$\sigma$} &  \multicolumn{1}{c}{$R^2$} &  \multicolumn{1}{c}{$\rho$} &  \multicolumn{1}{c}{$\sigma$} &  \multicolumn{1}{c}{$R^2$} &  \multicolumn{1}{c}{$\rho$} \\ \hline

$M_*$						& 0.19	& 0.91	& 0.95	& 0.21	& 0.89	& 0.94	& 0.23	& 0.84	& 0.92	\\ 
$Z_{\rm SFR}$					& 0.16	& 0.74	& 0.86	& 0.20	& 0.67	& 0.82	& 0.24	& 0.47	& 0.69	\\  
SFR							& 0.43	& 0.56	& 0.75	& 0.53	& 0.57	& 0.76	& 0.41	& 0.71	& 0.84	\\ 
$\rm M_{HI}$					& 0.31	& 0.36	& 0.59	& 0.24	& 0.60	& 0.77	& 0.24	& 0.71	& 0.84	\\ 
$\rm M_{H2}$					& 0.28	& 0.51	& 0.72	& 0.31	& 0.57	& 0.76	& 0.24	& 0.73	& 0.85	\\ \hline
$\rm M_{HI} | {\rm Baryon}$		& 0.25	& 0.59	& 0.77	& 0.21	& 0.69	& 0.83	& 0.21	& 0.78	& 0.88	\\ 
$\rm M_{H2} | {\rm Baryon}$		& 0.13	& 0.88	& 0.94	& 0.17	& 0.87	& 0.93	& 0.17	& 0.85	& 0.92	\\ \hline
\end{tabular}
\caption{\small{The scatter $\sigma$, the regression score $R^2$ and the Pearson correlation coefficient $\rho$ for the \mufasa\ vs. \galann\ scatter plots from Figs.~\ref{fig:only_z0}-\ref{fig:only_z1}.} The last two rows correspond to Fig.~\ref{fig:z0_stage2} where we additionally supply as inputs the baryonic properties $M_*$, SFR, and $Z$.}
\label{tab:SUMMARY_ALL}
\end{table*}

Comparing to the Illustris-based ML study of~\citet{Kamdar2016}, we find that our values of $R^2$ and $\rho$ for $M_*$ and SFR agree quite well with their predictions, showing that this level of agreement is generally independent of the details of ML algorithm and training simulation.  However, for gas mass and metallicity, our correlation measures are somewhat worse than their quoted values. The reason for the discrepancy is not clear, although it could stem from the way these quantities are defined and calculated in Illustris vs. \mufasa. It could also be because Illustris, having a larger dynamic range than \mufasa, contains many more galaxies, and thus is able to obtain a tighter fit.  Judging directly from their figures, it appears that the level of scatter between the true and predicted values for these quantities are broadly similar as ours.

In summary, our \galann\ framework is able to accurately recover the mean stellar, gas, and star formation contents of galaxies versus halo mass from \mufasa\ at all redshifts.  However, individual galaxies can deviate substantially, in a manner that depends on the quantity being examined; $M_*$ and $Z$ are most tightly predicted, while gas content and SFR are less well predicted at $z=0$.  The latter predictions improve significantly when going to higher redshifts.

\subsection{Second-parameter correlations}\label{sec:CORRELATIONS}

\begin{figure*}
     \includegraphics[scale=3]{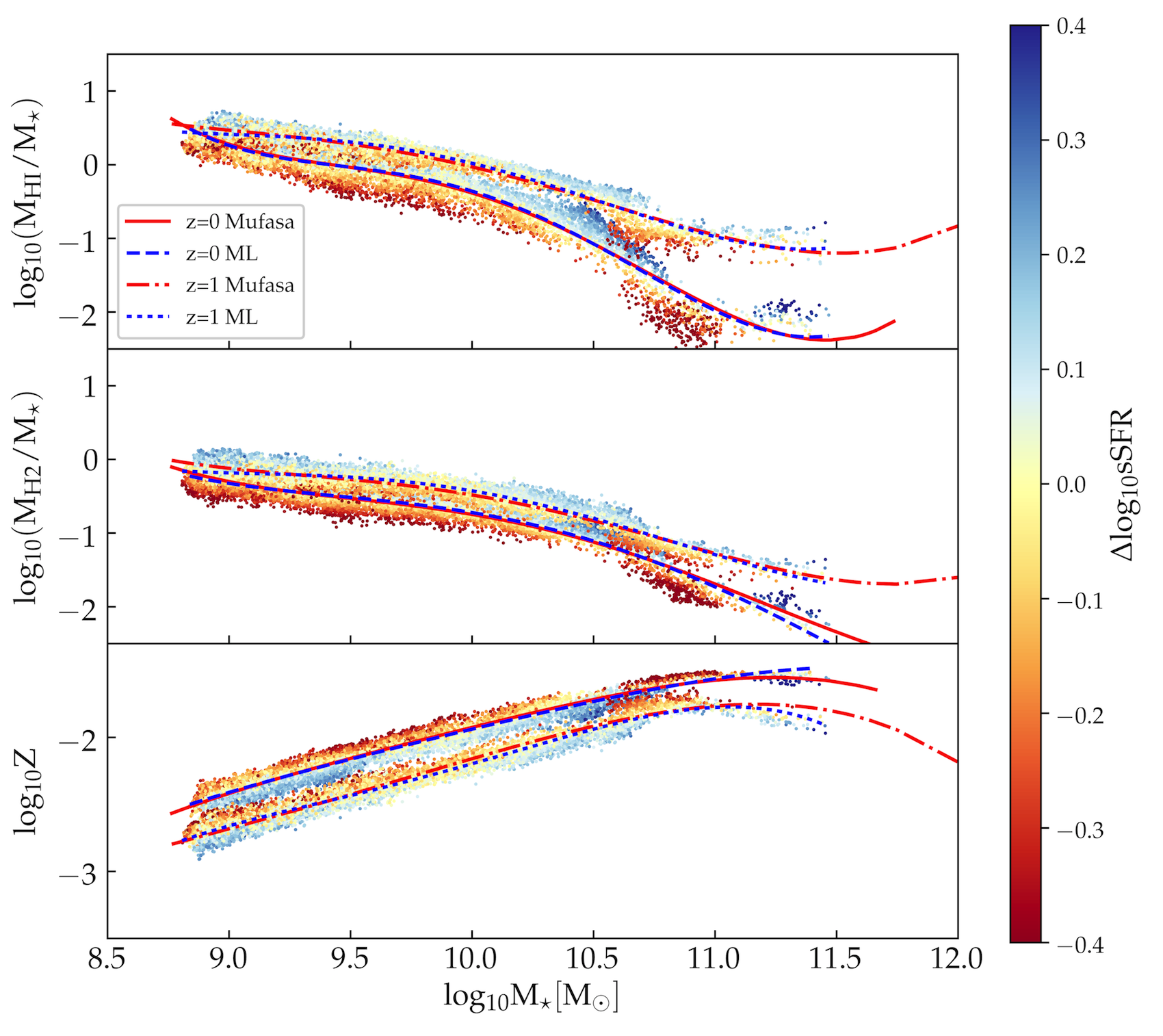}
        \caption{\small{Second-parameter correlations at $z=0$ and $z=1$. The panels (from top to bottom) show the \ion{H}{i} fraction ($\rm f_{HI}$), $\rm H_2$ fraction ($\rm f_{H2}$) and metallicity ($Z$) of galaxies as a function of their stellar mass $M_*$. The mean scaling relations are shown by the red (solid and dash-dotted, for \mufasa) and blue (dashed and dotted for \galann\ predictions) lines, and are in very good agreement with each other. Galaxies are color-coded by their distance from the mean $M_*$-sSFR relation ($\Delta\log_{10}$sSFR) obtained from \galann\ (blue dashed and dotted lines). At each redshift, for a given $M_*$, galaxies with higher sSFR than the mean relation tend to be metal-poor with a higher than average \ion{H}{i} and $\rm H_2$ content.}}
    \label{fig:correlations_z0}
\end{figure*}

\begin{figure*}
     \includegraphics[scale=3]{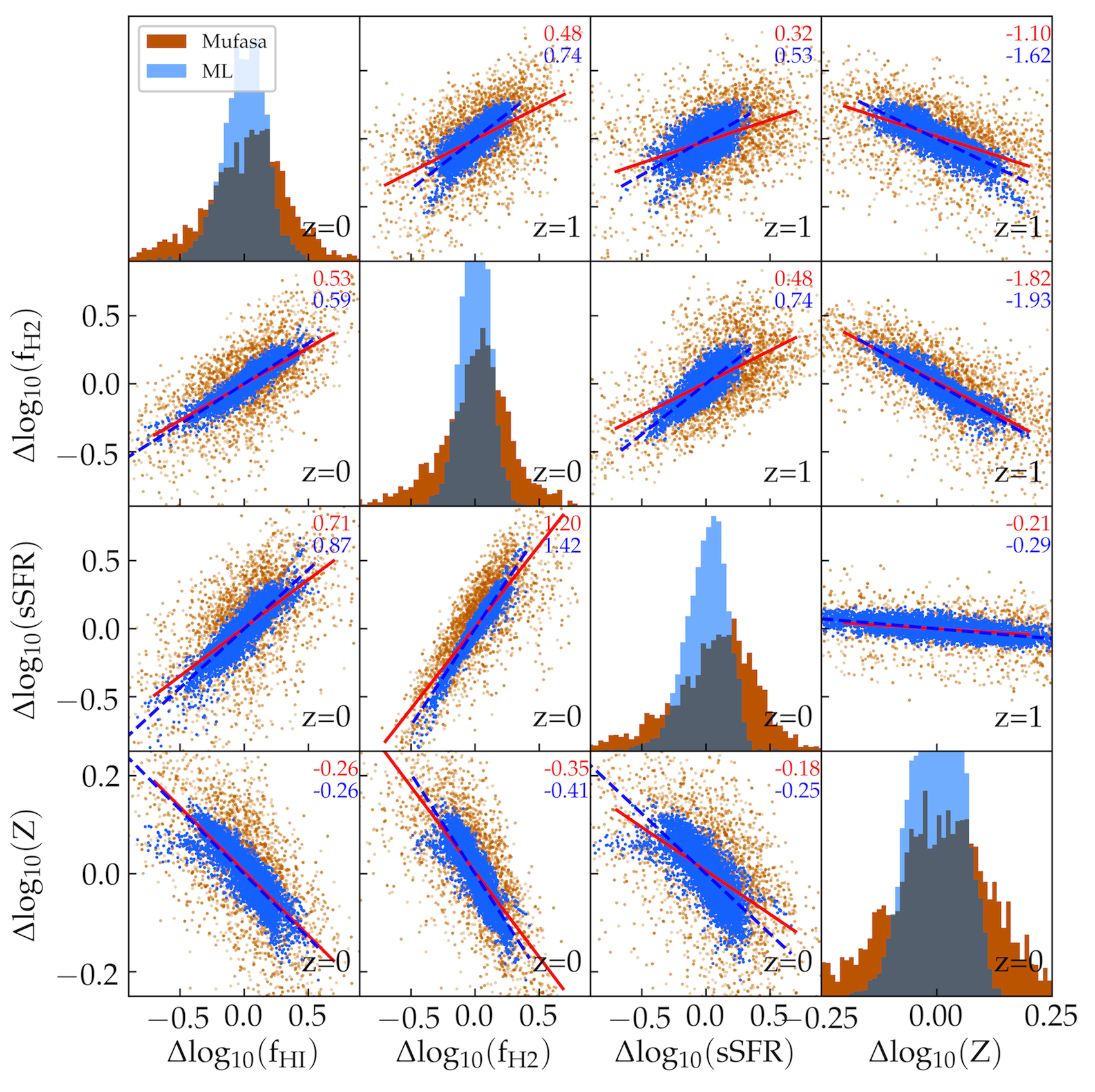}
        \caption{\small{Deviation of \ion{H}{i} fraction ($\rm f_{HI}$), $\rm H_2$ fraction ($\rm f_{H2}$) and metallicity ($Z$) of galaxies from their mean scaling relations versus $M_*$ (i.e. the red solid and blue dashed lines in Fig.~\ref{fig:correlations_z0}), at $z=0$ (lower left triangle) and $z=1$ (upper right triangle). The brown and blue points correspond to galaxies from the \mufasa\ simulation and \galann\ respectively. Histograms of each quantity at $z=0$ are shown along the diagonal.  The best-fit power laws (red solid line for \mufasa\ and blue dashed line for \galann) are overplotted, and their slopes are listed in the upper right of each panel.  Generally, the ML recovers the trend and approximate slope of the \mufasa\ deviation plots, showing that it is capable of learning higher-order correlations.}}
    \label{fig:correlations_diff_z0}
\end{figure*}

In the real Universe, it is known that the observed scatter in metallicity, gas content, and SFR are correlated.  For instance, it is observed that at a given stellar mass, a galaxy with high SFR will tend to have higher gas content~\citep{Bothwell13} and lower metallicity~\citep{Lara-Lopez10,Mannucci10}.  This can be understood theoretically in a model in which fluctuating gas inflow drives galaxies off the mean relation with respect to stellar mass, but subsequent processes tend to recover the galaxies back onto their equilibrium relations~\citep{Dave2011b,Dave2017a}.  Galaxies with temporarily elevated inflows (such as a merger) of lower-metallicity gas achieve higher \ion{H}{i} and $\rm H_2$ content, which stimulates SFR; conversely a lull in accretion results in more of a closed-box situation where the gas gets consumed and more metals are produced.  In the previous section we saw that \galann\ can well reproduce mean trends, but the individual scatter can be substantial, particularly for gas content and SFR.  To investigate this further, we explore whether \galann\ recovers the intrinsic second-parameter correlations in the scatter of these quantities.

Fig.~\ref{fig:correlations_z0} shows the second-parameter dependences with specific SFR (${\rm sSFR \equiv SFR}/M_*$) for the RF predictions at $z=0$ and $z=1$. The panels (from top to bottom) show the \ion{H}{i} fraction ($\rm f_{HI}$), $\rm H_2$ fraction ($\rm f_{H2}$) and metallicity ($Z$) as a function of galaxy stellar mass $M_*$. Galaxies are color-coded by their distance from the mean $M_*$-sSFR relation obtained from \galann\ (blue dashed and dotted lines), such that galaxies with above average sSFR are bluer, and galaxies with lower sSFR are redder. These can be compared e.g. to plots such as Fig. 3 and Fig. 5 in~\citet{Dave2017a}.  Like in the case versus halo mass, the mean relations versus galaxy stellar mass agree extremely well between \mufasa\ and \galann\ at both $z=0$ and $z=1$.

More remarkable is the fact that \galann\ recovers the second-parameter correlations in the scatter, despite not being provided any baryonic information.  That is, for a given $M_*$, galaxies with higher sSFR than typical are also metal-poor with an above-average \ion{H}{i} and $\rm H_2$ fractions.  The correlations also exist at redshifts $z=2$ (not shown), although are somewhat less evident.  These trends qualitatively mimic those seen in observations~\citep{Salim2014,Salim2015} and hydrodynamic simulations~\citep{Dave2011b}.

We can conduct a more quantitative comparison to \mufasa\ using deviation plots~\citep{Dave2017a}, in which we show the deviation of the quantities from their mean scaling relations vs $M_*$, plotted against each other.  Fig.~\ref{fig:correlations_diff_z0}, shows the same data in Fig.~\ref{fig:correlations_z0}, re-plotted in terms of deviations of $\rm f_{HI}$, $\rm f_{H2}$, sSFR and $Z$ from their respective mean scaling relations vs. $M_*$. Brown points show the second-parameter dependences directly from the \mufasa\ simulation, while the blue points show the \galann\ predictions.  Lines (red solid for \mufasa\ and blue dashed for \galann) show best-fit power laws, with the slopes of the trends listed in the upper right of each panel. 

The predictions from \galann\ reasonably well reproduce \mufasa\ power law slopes of all these deviations, though not perfectly so.  The broad agreement is nonetheless a remarkably stringent test of \galann's ability to mimic simulated galaxy properties in terms of higher-order correlations.  The scatter in the deviations is also reasonably reproduced, but as with the overall scatter in the quantities themselves, the blue \galann\ points do not quite span the full range of \mufasa's red points, reflecting the the smaller scatter predicted by \galann\ than seen in \mufasa.

More broadly, the second-parameter dependence shows the power of machine learning in being able to be trained on all the available information to predict galaxy properties in greater detail. A simple parameterization of the mean scaling relations of individual properties versus halo mass would not contain such second-parameter information; this is obtained purely because the ML has been trained on the full set of galaxy properties in the simulation, which contain these correlations. This hints that providing some baryonic information, particularly related to star formation rate, may substantially improve the ML predictions.  We examine this later on.

\begin{figure}
     \includegraphics[width=\columnwidth]{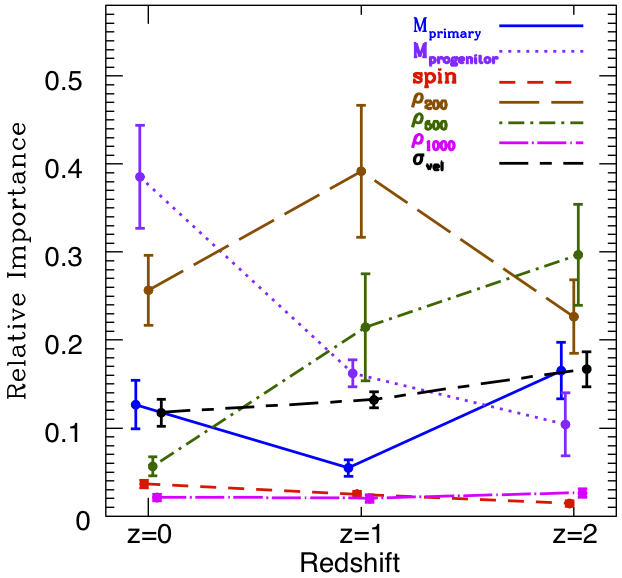}
        \caption{\small{Relative importance of input features as determined in RF from $z=0$ to $z=2$. The error bars correspond to the standard deviation of each quantity from 10-fold cross validation. See \S\ref{sec:FEATURE_IMPORTANCE} for exact definitions of these quantities.}}
    \label{fig:relative_importance}
\end{figure}

\begin{figure}
     \includegraphics[width=\columnwidth]{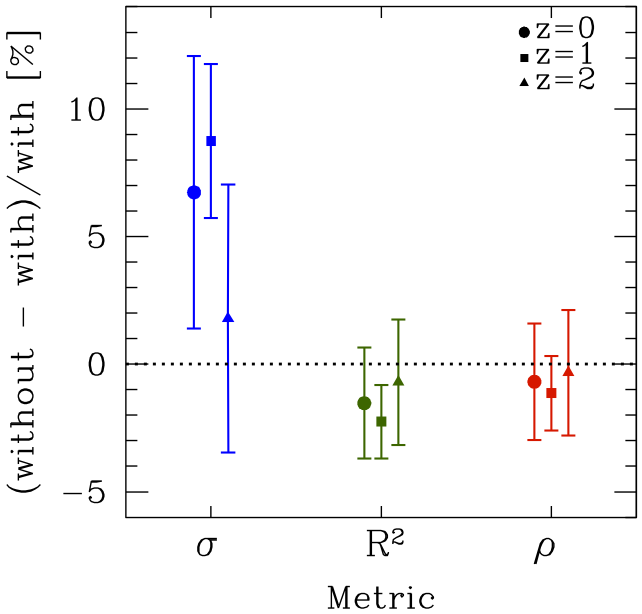}
        \caption{\small{Percentage change in the three metrics $\sigma$, $R^2$ and $\rho$ for $M_*$ when merger history of DM halos is neglected during the ML training phase. The error bars correspond to the standard deviation from 10-fold cross validation. We deduce that neglecting halo merger history only slightly impacts the overall fits. This plot employs to RF; similar results are obtained for other ML methods.}}
    \label{fig:progenitor_summary}
\end{figure}

\subsection{Relative importance of inputs}\label{sec:FEATURE_IMPORTANCE}

The random forest algorithm uses a decision tree to generate the mapping from input to output quantities.  For RF (as well as DT and ET) an input feature is ranked in importance according to the depth of the decision tree to which the feature in question is used in decision making.  This thus provides a way to estimate, at least for these algorithms, which input variables are most responsible for governing the mapping from inputs to outputs.  We examine these relative importances for our RF results in this section.

Fig.~\ref{fig:relative_importance} shows the relative importance of the input features at $z=0-2$.  The input features are as described in \S\ref{sec:ML_overview}, with two new aggregate definitions for the primary and progenitor halo masses: $\rm M_{primary}$ corresponds to the sum of importances for the current redshift and its immediate predecessor, while $\rm M_{progenitor}$ is the combined importance of the remaining earlier progenitors.  We find these aggregate quantities are more stable and useful than considering each individual progenitor mass separately.  The error bars reflect the variance over the 10 ensemble ML trials.

At $z=0$, the progenitor mass and nearby (200~kpc) environment are the driving inputs, and combine to produce two-thirds of the total importance.  Primary mass and velocity dispersion are also significant drivers at the $\sim10$\% level each.  At face value, this suggests that including the immediate mass growth (i.e. merger) history of dark matter halos is crucial for obtaining a good ML mapping.  

Going to higher redshifts, the intermediate-scale (500~kpc) environment becomes much more important, while the progenitor mass commensurately reduces in importance.  Note that these environmental measures are in comoving kpc, so in fact at $z\sim1-2$, this corresponds to $\sim 200$ physical kpc.  At face value, this suggests that galaxies properties at all epochs are mostly driven by environmental processes on a $\sim 200$~kpc physical scale, although one must be careful not to read too much physics into these importances.  Also, the halo velocity dispersion is reasonably important, although in simulations this is very tightly correlated with the halo mass, so it is unclear whether this is providing distinct constraints.

Another interesting result is that at all epochs, the large-scale (Mpc) environment and spin are unimportant.  The spin in particular may be surprising, given that simple galaxy models would associate spin as a determining factor for whether a galaxy collapses early and forms a massive elliptical (low spin) or it is a disky star-forming object until today (high spin).  At least within this ML framework, it appears the halo spin is not an important predictor of galaxy quantities such as mass and SFR.

Given the importance of the progenitor mass in RF, we explored more directly the results if we removed the progenitor information from the inputs.  For this, we retrain and rerun the RF network with the only halo mass information provided being the current mass, without any progenitor masses.  This is expected to give a worse fit.  We quantify how much worse in Fig.~\ref{fig:progenitor_summary}, which shows the percentage change in the scatter and correlation properties of the $M_*$ distribution when halo merger history is neglected.  A worse fit results in a larger $\sigma$, and a smaller $R^2$ and $\rho$.  Despite the RF importance level, the change is relatively minor, with an increase in $\sigma$ at the few to 10\% level, and only a percent-level reduction in $R^2$ or $\rho$.  This suggests that the progenitor information is actually sufficiently encoded in the remaining quantities that removing this quantity does not greatly impact the overall fits.  This cautions against reading too much into the details of importance levels from the RF algorithm.

\subsection{Improvement with baryonic inputs}\label{sec:BARYONIC_INPUT}

N-body simulation only contain information about dark matter halos.  However, it is possible to populate N-body simulations with galaxies using simplified prescriptions.  A traditional approach is to use SAMs, but these typically contain a large number of free parameters.  A more recent approach is to use so-called equilibrium~\citep{Dave2012} or bathtub-type models~\citep{Bouche2010,Lilly2013}.  In such models, galaxy evolution is fully captured via a relatively small number of physical parameters that quantify the baryon cycling process, which can be constrained to observations with well-characterized Bayesian posteriors.  For instance, \citet{Mitra2015} found that with 8 free parameters that describe ejective feedback, preventive feedback, and wind recycling, it was possible to well predict the evolution of the stellar mass--halo mass, star formation rate--stellar mass, and mass-metallicity relations from $z=0-2$.  The outputs of such a model are, for a given halo with a mass assembly history, the resulting stellar mass, star formation rate, and metallicity of the halo's central galaxy.  Hence by running such a model on a halo merger tree, it becomes possible to predict $M_*$, SFR, and $Z$ for each halo (along with their histories), which can then be used as an input to \galann\ rather than as an output.  In the present case, the remaining outputs are then the \ion{H}{i} and $\rm H_2$ masses.  

\begin{figure*}
     \includegraphics[width=\columnwidth]{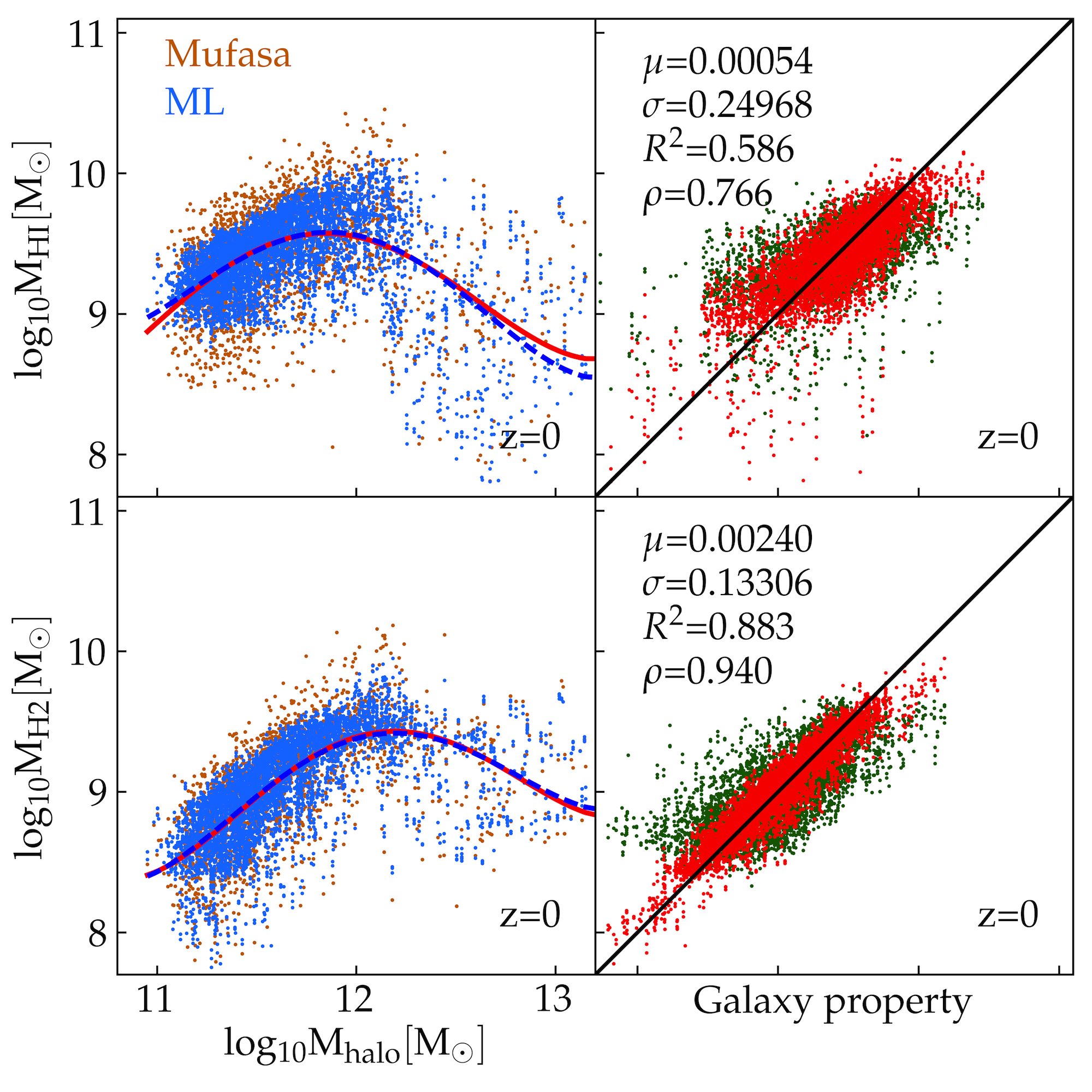}
     \includegraphics[width=\columnwidth]{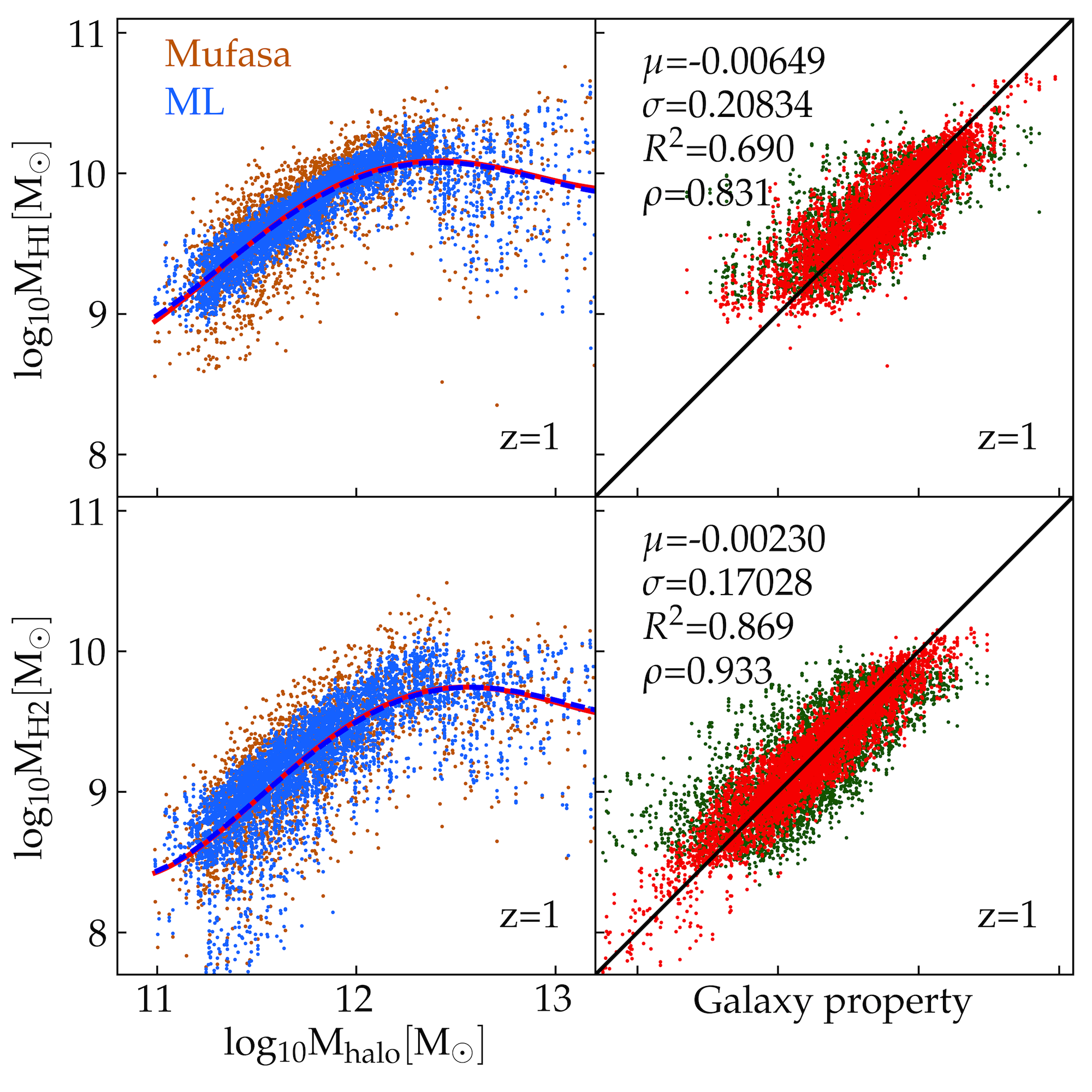}
        \caption{\small{Galaxy \ion{H}{i} (top panels) and $\rm H_2$ (bottom panels) mass vs. host halo mass, at $z=0$ (left 4 panels) and $z=1$ (right 4 panels). In making the predictions, RF was trained using the usual halo quantities, as well as galaxy quantities $M_*$, SFR and $Z$. Inclusion of baryonic information of galaxies improves the \ion{H}{i} and $\rm H_2$ predictions considerably (red points). For comparison, see the \ion{H}{i} and $\rm H_2$ predictions in Fig.~\ref{fig:only_z0} and Fig.~\ref{fig:only_z1} (reproduced here by green points).}}
    \label{fig:z0_stage2}
\end{figure*}

In this section, we examine how much improvement one can expect in $\rm M_{HI}$ and $\rm M_{H2}$ by providing the equilibrium model output quantities as {\it inputs} to \galann, in addition to the full complement of dark matter properties.  Fig.~\ref{fig:z0_stage2} shows the result, at $z=0$ and $z=1$.  Comparing to the original determinations in Figs.~\ref{fig:only_z0} and \ref{fig:only_z1}, it is clear that the scatter $\sigma$ is significantly reduced, and $R^2$ and $\rho$ are higher.  It is still not ideal, but clearly provides a step forward towards accurate predictions of the gas content of galaxies.  

\begin{figure}
     \includegraphics[width=\columnwidth]{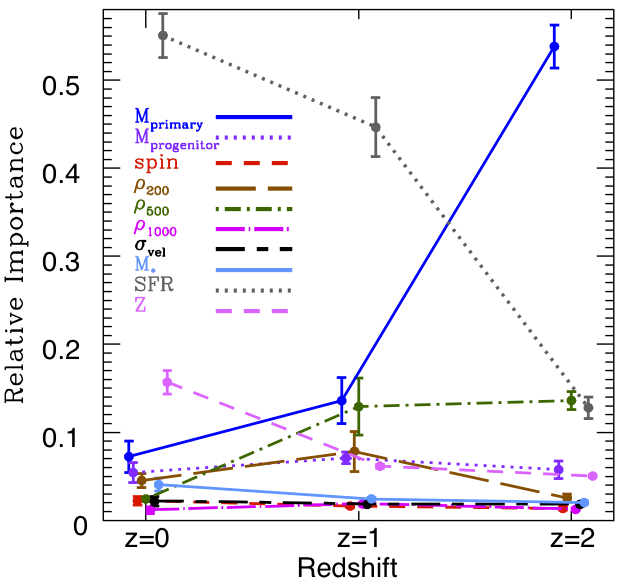}
        \caption{\small{Relative importance of input features as a function of redshifts as in Fig.~\ref{fig:relative_importance}, now including the baryonic inputs.  The dramatic improvement in predictions is driven primarily by including information about the current SFR (gray dotted line) of the galaxy; stellar mass and metallicity are important as well but clearly sub-dominant to SFR.}}
    \label{fig:relative_importance_stage2}
\end{figure}

To understand which baryonic quantity provided this improvement, in Fig.~\ref{fig:relative_importance_stage2}, we show the relative importance of the input features when galaxy quantities $M_*$, SFR and $Z$ are included on the input side of the network.  It appears that the key new quantity that is most responsible for the improvement is the SFR, at least at $z=0,1$; at $z=2$, the primary halo mass remains the most important.  This is not surprising since the gas content is most closely related to the SFR; the metallicity and halo mass are also helpful, and in fact at $z=0$ those quantities are the top 3 in importance. This experiment thus demonstrates that including some baryonic information as might be available from well-constrained analytic models can help to improve the gas predictions from \galann\ particularly for \ion{H}{i}.

\section{Comparing machine learning methods}\label{sec:COMPARE_ML}

So far we have focused on the RF algorithm for ML.  In this section we show the tests that led to this decision, by comparing various ML methods with regards to our correlation scores.

\begin{figure}
     \includegraphics[width=\columnwidth]{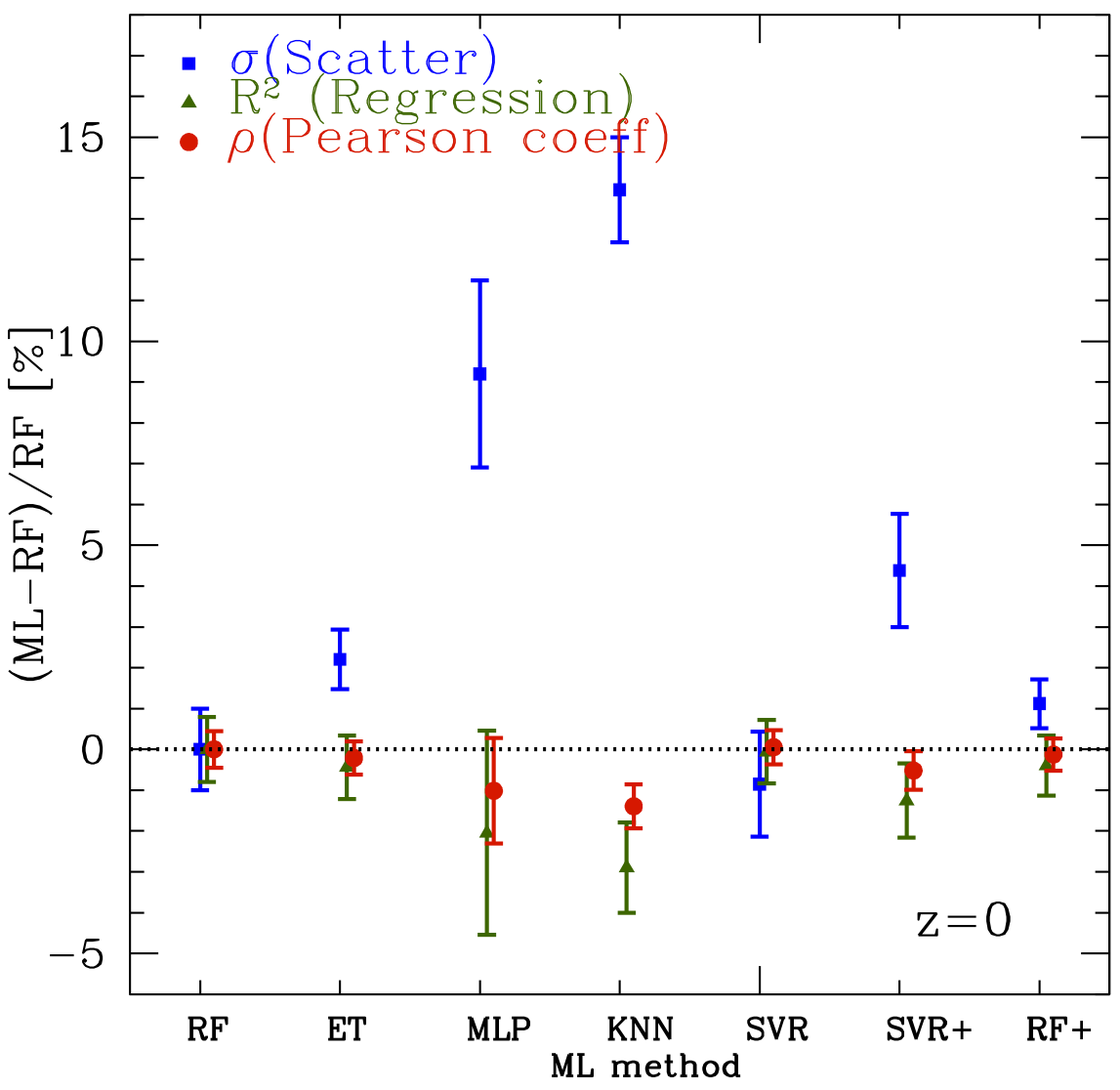}
        \caption{\small{The performance metrics (scatter $\sigma$, regression $R^2$, and Pearson correlation coefficient $\rho$), quoted in Fig.~\ref{fig:only_z0} at $z=0$ for galaxy stellar mass $M_*$, are plotted for the ML algorithms explored in this paper. The results are shown relative to the metric values for RF. For each algorithm, the error bars correspond to the standard deviation of each quantity from 10-fold cross validation. SVR+ and RF+ correspond to the boosted versions of SVR and RF using AdaBoostRegressor. All ML routines have comparable regression ($R^2$, green) and Pearson ($\rho$, red) coefficients. Compared to RF, the scatter ($\sigma$, blue) is $\sim 10-15\%$ larger for MLP and kNN. The corresponding plots for the other galaxy quantities (${\rm M_{HI}, M_{H2}, SFR,} Z$), and for quantities predicted at $z=1$ and $z=2$, are similar. We deduce that for the most part the different algorithms all deliver similar performance, at least for $R^2$ and $\rho$. }}
    \label{fig:ML_summary_z0}
\end{figure}

Fig.~\ref{fig:ML_summary_z0} shows the fractional difference in the scatter ($\sigma$, blue), regression ($R^2$, green), and Pearson correlation coefficient ($\rho$, red), for our tested ML methods, normalized to the RF case.  The leftmost points correspond to RF, and are by definition unity.  Compared to RF, ET and SVR perform approximately as well, SVR perhaps even slightly better. MLP and kNN are significantly worse, with higher scatter and lower correlation coefficients. All three metrics degrade slightly when RF and SVR are combined with AdaBoostRegressor (denoted as RF+ and SVR+), possibly due to over-fitting.  Owing to this analysis, we chose to use Random Forests to quote our ML results.

To further explore ways to improve the method, we tried combining the results from our full ensemble of ML methods, either by taking the mean result among all the methods for each galaxy, or taking the median.  Unfortunately, in neither case was there any noticeable improvement in the predictions.  The issue likely traces back to the scatter.  As shown in Fig.~\ref{fig:only_z0} and related figures, the scatter in \galann\ predictions around the mean relations is too small compared to the true range.  All other ML methods behave comparably -- none are able to properly recover the full scatter in the relations.  It appears all ML methods are like RF in their tendency to push the predictions towards the mean relations, and avoid outliers.  It remains to be seen if there is an ML algorithm that is better suited to reproducing the full distribution of the training set rather than biasing the predictions towards the mean.

\section{Summary and Discussion}\label{sec:DISCUSS}

We have investigated machine learning as a framework to populate baryonic galaxies inside dark matter halos.  We train on the \mufasa\ cosmological hydrodynamic simulations, and test the efficacy of various machine learning regression algorithms against a distinct subset of galaxies from the same simulation. Dark matter inputs to the machine learning are the halo masses over a range of cosmic ages, the halo densities in radii of 200, 500 and 1000 kpc, the halo spin and the halo velocity dispersion. The predicted outputs are stellar mass, \ion{H}{i} and $\rm H_2$ mass, star formation rate (SFR) and SFR-weighted metallicity both for the entire halo and the central galaxy. 

Our main conclusion is that machine learning is able to correctly predict the mean baryonic properties well at all halo masses, except at the low mass end where they are over-estimated which is likely an artifact of limited dynamic range. Somewhat of a surprise was the result that including halo merger history leads to only marginally improved results.  

We find that none of the algorithms capture the full range of observed scatter in baryonic properties. As a result, ensembling the outputs from the various algorithms does not lead to improved results. The machine learning predictions do however capture the second-parameter correlations implicit in the \mufasa\ simulations which encode important physics, e.g. high SFR is linked to higher gas content and lower metallicity.   

We show that if some baryonic information is available through additional pre-modeling such as provided by the equilibrium model, the machine learning algorithms can exploit this and can match the observed scatter much better, giving improvements of up to $100\%$, especially in $\rm H_2$ predictions.  When including the baryonic information the most important feature in the machine learning fits is the star formation rate. 

Finally, we compare seven common regression algorithms (and EDEN -- a bespoke evolutionary deep learning algorithm, but not shown in this paper). We also explore an ensemble approach looking at the mean and median of the outputs of the various algorithms. Performance is broadly similar, with the random forests and support vector regression marginally better than others.  This suggests that the limitations are intrinsic to the amount of information contained in the dark matter distribution, rather than in the performance of the ML algorithm.

The ML approach to populating galaxies into halos has intrinsic limitations, some of which can be at least partially mitigated.  For one, any ML framework will only be as good as the training data used.  In our case, we have used the \mufasa\ simulation, which provides quite good agreement with a wide range of (though not all) galaxy properties~\citep{Dave2016,Dave2017a,Dave2017b,Rafieferantsoa2018}; other modern simulations can achieve similar agreement albeit with varying input physics~\citep{Somerville2015}.  As simulations continue to improve in realism, applying this ML framework to the latest galaxy formation models will provide increasingly robust predictive power.

A second limitation is that the dynamic range in the hydrodynamic simulation will naturally be  smaller than that of the target N-body simulation (else there would be no reason to do the ML).  As such, populating large-volume simulations will require an extrapolation beyond the training set.  One way to mitigate this would be to run zoom simulations targeting dwarfs and galaxy clusters~\citep{Cui2016} using the exact same physical model, thereby providing anchor points at large and small halo masses for training.

Another limitation is that we are training on halos and galaxies taken from a hydro simulations, but the gas dynamics could potentially have a non-negligible impact on halo properties.  A way to mitigate this would be to run a pure dark matter version of the same hydro simulation, and then match up halos individually, and use the halo properties from the N-body run rather than the hydro run.  This would more accurately mimic the situation in a pure dark matter run.  Indeed, the ML framework could even calibrate how halos are modified owing to baryonic processes, which is interesting in of itself.

Finally, we have concentrated here on central galaxies, whereas fully populating an N-body run will require including the satellites as well.  This is perhaps the most challenging aspect, because it will require identifying subhalos in N-body simulations as sites to populate with baryonic satellites, but sub-halo identification and numerical resolution issues in N-body simulations can be severe~\citep{vandenBosch2018}.  An alternative approach would be to instead use the merger tree from the N-body simulation, and follow the satellites using a calibrated analytic model that does not suffer from numerical convergence issues~\citep{Simha2017}.

Modulo the above caveats that we aim to tackle in future work, our results suggest that machine learning is a promising approach to populate galaxies into halos in large N-body simulations, which is useful for a wide variety of galaxy formation and cosmology applications.  One such application is \ion{H}{i} intensity mapping as will be done with upcoming surveys such as CHIME, HIRAX, and MeerKLASS on MeerKAT.  For this, the ML would need to output the halo \ion{H}{i} content which is straightforward, and this would be mapped into halos from a dark matter simulation.  Another application would be weak lensing galaxy shape correlations, for which the ML would need to be trained on galaxy ellipticities which would again be straightforward within our framework although we have not done it here. Future work includes identifying dark matter features that can act as suitable proxies for baryonic information, tuning more sophisticated ML algorithms such as deep learning, using recurrent neural networks to learn the temporal evolution of halos, applying to larger simulations with more galaxies to avoid being limited by the size of training samples, and marrying it with simple analytic models to additionally input an estimate of the instantaneous SFR.  The flexibility of ML and the ability to assess its accuracy based on testing sets suggests that it is a worthwhile alternative approach to connecting galaxy baryonic properties to dark matter halos within large-scale structure.

\section*{Acknowledgements}

The authors thank the referee for providing constructive comments and help in improving the contents of this paper. SA acknowledges Laboratoire Univers et Th\'eories (LUTh) at the Observatoire de Paris, for allowing its computing resources be used for running {\sc Scikit-learn} machine learning library.  RD acknowledges support from the South African Research Chairs Initiative and the South African National Research Foundation.  RD further acknowledges long-term visitor support provided by the Simons Foundation's Center for Computational Astrophysics and the Distinguished Visitor Program at Space Telescope Science Institute.  The \mufasa\ simulations were run on the Pumbaa astrophysics computing cluster hosted at the University of the Western Cape, which was generously funded by UWC's Office of the Deputy Vice Chancellor, and were run with revision e77f814 of {\sc Gizmo} hosted at {\tt https://bitbucket.org/rthompson/gizmo}. 
\bibliographystyle{mnras}
\bibliography{galann}

\renewcommand\thesection{\Alph{section}}
\setcounter{section}{0}
\renewcommand\thesubsection{\thesection\arabic{subsection}}

\end{document}